\documentclass[12pt,preprint]{aastex}

\shorttitle{Simplified supernova explosions}
\shortauthors{Cardall and Budiardja}

\usepackage[pdftex,plainpages=false,pdfpagelabels=true,breaklinks=true,pagebackref=true]{hyperref}

\begin{document}


\title{Stochasticity and efficiency of simplified core-collapse supernova explosions}


\author{Christian Y. Cardall}
\affil{Physics Division, Oak Ridge National Laboratory, Oak Ridge, Tennessee 37831-6354, USA\altaffilmark{1}}
\email{cardallcy@ornl.gov}

\author{Reuben D. Budiardja}
\affil{National Institute for Computational Sciences, University of Tennessee, Knoxville, TN 37996, USA}
\email{reubendb@utk.edu}


\altaffiltext{1}{Department of Physics and Astronomy, University of Tennessee, Knoxville, Tennessee 37996-1200, USA}


\begin{abstract} 
We present an initial report on 160 simulations of a highly simplified model of the post-bounce core-collapse supernova environment in three spatial dimensions (3D).
We set different values of a parameter characterizing the impact of nuclear dissociation at the stalled shock in order to regulate the post-shock fluid velocity, thereby determining the relative importance of convection and the stationary accretion shock instability (SASI).
While our convection-dominated runs comport with the paradigmatic notion of a `critical neutrino luminosity' for explosion at a given mass accretion rate (albeit with a nontrivial spread in explosion times just above threshold), the outcomes of our SASI-dominated runs are much more stochastic: a sharp threshold critical luminosity is `smeared out' into a rising probability of explosion over a $\sim 20\%$ range of luminosity.
We also find that the SASI-dominated models are able to explode with 3 to 4 times less efficient neutrino heating, indicating that progenitor properties, and fluid and neutrino microphysics, conducive to the SASI would make the neutrino-driven explosion mechanism more robust.
\end{abstract}



\section{Introduction}
\label{sec:Introduction}

Despite a notorious lack of robustness, the delayed neutrino-driven mechanism has persisted for 30 years as the leading paradigm of typical core-collapse supernova explosions 
\citep{Bethe1985Revival-of-a-st,Kotake2012Multimessengers,Janka2012Explosion-Mecha,Burrows2012Perspectives-on}.
Gravitational collapse of a massive ($\gtrsim~8\ M_\odot$) star halts with a `bounce' at nuclear density, launching a shock wave that stalls at $100-200\ \mathrm{km}$, a few times the radius of the hot nascent neutron star.
Neutrino fluxes from the neutron star heat the outer post-shock region.
In spherically symmetric distillations of the problem, if the neutrino luminosity is sufficiently high for a given mass accretion rate, the ram pressure of infalling matter is overcome and the shock resumes its outward journey \citep{Burrows1993A-Theory-of-Sup,Janka2001Conditions-for,Yamasaki2007Stability-of-Ac,Pejcha2012The-Physics-of-,Fernandez2012Hydrodynamics-o,Muller2015Non-radial-inst,Murphy2015An-Integral-Con}.
Multidimensional fluid instabilities increase the dwell time of matter in the heating region and the average stalled shock radius relative to spherical symmetry; these effects are often characterized as reducing this critical or threshold luminosity for explosion \citep{Burrows1995On-the-Nature-o,Janka1996Neutrino-heatin,Yamasaki2005Effects-of-Rota,Ohnishi2006Numerical-Analy,Yamasaki2006Standing-Accret,Iwakami2008Three-Dimension,Murphy2008Criteria-for-Co,Fernandez2009Dynamics-of-a-S,Hanke2012Is-Strong-SASI-,Dolence2013Dimensional-Dep,Couch2013On-the-Impact-o,Couch2015The-Role-of-Tur,Muller2015Non-radial-inst,Fernandez2015Three-dimension}.
Beyond these generally more favorable conditions, 
it seems that the runaway expansion of one or a few large, persistent, buoyant bubbles, inflated by neutrino heating, is the final proximate agent propelling the shock on its way \citep{Herant1995The-convective-,Thompson2000Accretional-Hea,Fernandez2009Dynamics-of-a-S,Dolence2013Dimensional-Dep,Wongwathanarat2013Three-dimension,Couch2013On-the-Impact-o,Handy2014Toward-Connecti,Fernandez2014Characterizing-,Muller2015Non-radial-inst,Fernandez2015Three-dimension,Lentz2015Three-dimension}.
Here we consider two questions related to the delayed neutrino-driven mechanism.

First, does this system exhibit qualitatively significant sensitive dependence on initial conditions? 
How reproducible and representative is a single simulation, compared to others with nearly identical initial states? 
The concern is that 3D simulations with more sophisticated neutrino transport take tens of millions of processor hours and months to run; only a handful have been reported \citep{Hanke2013SASI-Activity-i,Tamborra2014Self-sustained-,Melson2015Neutrino-driven-I,Melson2015Neutrino-driven-II,Lentz2015Three-dimension}.

Second, might the stationary accretion instability (SASI) materially contribute to supernova explosions \citep{Hanke2012Is-Strong-SASI-,Muller2012New-Two-dimensi,Hanke2013SASI-Activity-i,Fernandez2014Characterizing-,Tamborra2014Self-sustained-,Iwakami2014Parametric-Stud,Melson2015Neutrino-driven-II,Fernandez2015Three-dimension}; or is it generically subdominant to an extent that convection is all that matters \citep{Burrows2012An-Investigatio,Murphy2013The-Dominance-o,Dolence2013Dimensional-Dep,Handy2014Toward-Connecti,Takiwaki2014A-Comparison-of,Couch2014High-resolution,Abdikamalov2015Neutrino-driven,Melson2015Neutrino-driven-I,Lentz2015Three-dimension}?
In the absence of heating, a stationary accretion shock does not remain stationary: small asphericities result in large-scale (multipole $\ell = 1, 2$), nonlinear, oscillatory motions---typically, spiral waves in 3D \citep{Blondin2003Stability-of-St,Blondin2007Pulsar-spins-fr,Fernandez2010The-Spiral-Mode,Guilet2012On-the-linear-g}.
This increases the average shock radius, but not to explosion;
neutrino heating, which drives convection and inflates large buoyant bubbles, is ultimately required.
But strong neutrino heating tends to suppress the SASI, as its persistent larger-scale bubbles interrupt cyclic, globally coherent SASI flows \citep{Burrows2012An-Investigatio,Fernandez2014Characterizing-,Couch2014High-resolution,Iwakami2014Parametric-Stud,Fernandez2015Three-dimension}.
The question is whether the SASI facilitates the heating and inflation of large buoyant bubbles to any important extent, by lengthening the advection timescale and generating entropy gradients \citep{Scheck2008Multidimensiona}.

\section{Model}

We explore these questions with a simplified model of the post-bounce supernova environment, closely following \citet{Fernandez2014Characterizing-} and \citet{Fernandez2015Three-dimension}, allowing for many realizations and control over the nature of the dominant instability.

We obtain initial conditions from stationary, spherically symmetric, and nonrelativistic fluid equations for baryon, momentum, and energy conservation. 
We use an ideal gas equation of state with adiabatic index $\gamma=4/3$,
central gravity with fixed point mass $M = 1.3~M_\odot$, and constant mass accretion rate $\dot M = 0.3~M_\odot~s^{-1}$.
A parameter $\varepsilon$ mimicks nuclear dissociation in the shock jump conditions.
We use the energy source
\begin{equation}
Q_\nu = \left(\frac{B}{r^2} - A\, p^{3/2}\right) \rho\; \Theta(\mathcal{M}_0-\mathcal{M}),
\label{eq:HeatingCooling}
\end{equation}
where $B$ and $A$ parametrize the magnitude of neutrino heating and cooling, $r$ is the radius, $p$ is the pressure, $\rho$ is the mass density, $\mathcal{M}$ is the Mach number, and $\mathcal{M}_0 = 2$ is a reference value separating the pre- and post-shock regions; 
the Heaviside step function $\Theta$ restricts heating and cooling to the post-shock region.

Initial conditions in the pre- and post-shock regions are joined by the shock jump conditions.
An inner reflecting boundary at $r_\mathrm{NS} = 40~\mathrm{km}$ represents the (fixed) surface of the hot nascent neutron star.
We choose a shock radius $r_{S0} =100~\mathrm{km}$ in the absence of neutrino heating ($B=0$).
A vanishing Bernoulli parameter and $\mathcal{M} = 6$ at $r_{S0}$ determine the nearly free-fall pre-shock solution. 
The $\varepsilon$-dependent shock jump conditions determine the initial conditions of integration from the shock to the inner boundary.
All simulations with a given $\varepsilon$ use the same $A$.
We determine these $A$ values empirically, as those which yield zero velocity at $r_\mathrm{NS}$ with $B=0$.
In setting up initial profiles for increasing values of $B$ (with $\varepsilon$ and $A$ fixed), the shock radius $r_S > r_{S0}$ is empirically determined as that which again yields zero velocity at $r_\mathrm{NS}$.

We evolve the system with \textsc{GenASiS} \citep{Cardall2014GENASIS:-Genera,Cardall2015GenASiS-Basics:}, using a multilevel mesh in Cartesian coordinates.
The coarsest level covers  $[-640\,\mathrm{km}, +640\,\mathrm{km}]$ in 3D. 
Three additional levels consist of nested spheres of radius $320\,\mathrm{km}$, $160\,\mathrm{km}$, and $80\,\mathrm{km}$.
The coarsest contains $128^3$ cells (cell width $10\,\mathrm{km}$);
successive levels are refined by a factor of 2 (finest level cell width $1.25\,\mathrm{km}$).
We interpolate the spherically symmetric initial conditions to the 3D multilevel mesh and add random $0.1\%$ pressure perturbations.
We implement the mock-up of nuclear dissociation with an additional continuity equation for the density of `dissociated baryons,' with appropriate sources for this and the energy equation.

\section{Results}

We present `convection-dominated' (C-series) and `SASI-dominated' (S-series) runs.
Convection only develops from small perturbations if the post-shock infall velocity is sufficiently slow; otherwise, incipient buoyant perturbations are advected to the neutron star before they can appreciably grow. 
In that case, the SASI is more free to operate. 
The Foglizzo number $\chi$---the ratio of advection time to buoyancy timescale in the region of net heating beneath the shock---is a quantitative measure of this, with $\chi \gtrsim 3$ allowing convection to develop from linear perturbations \citep{Foglizzo2006Neutrino-driven}.

Our C-series and S-series differ in dissociation parameter $\varepsilon$, which determines their convection-dominated vs. SASI-dominated character. 
By removing energy from the fluid, larger values increase the shock compression ratio and reduce the post-shock infall speed; 
this increases the advection time scale, and therefore $\chi$.
The values of $\chi$ in Table~\ref{tab:SeriesParameters} for the initial profiles used in the C-series and S-series are respectively well above and below the value $\chi = 3$ separating the convection-dominated and SASI-dominated regimes. 
The perhaps extreme $\varepsilon$ values serve to separate these regimes; the messier truth may lie between.

\begin{table}
\begin{center}
\caption{Simulation summary.\label{tab:SeriesParameters}}
\begin{tabular}{lcccccrcc}
\tableline\tableline
Dominant & $\varepsilon$ & $A$ & $B$ & $r_S$ & $\chi$ & $N_\mathrm{400}$ & $t_\mathrm{400,min}$ & ${\Delta t}_\mathrm{400}$ \\
Instability & $(\varepsilon_0)$ & $(A_0)$ & $(B_0)$ & $(r_{S0})$ &  & $ $ & (s) & (s) \\
\tableline
Convection & 0.3 & 0.0131 & 0.700 & 1.202 & 6.27 & 0 & \dots & \dots \\
(C-Series) &  & & 0.725 & 1.212 & 6.62 & 0  & \dots & \dots \\
& & & 0.750 & 1.223 & 6.97 & 0 & \dots & \dots \\
& & & 0.775 & 1.234 & 7.33 & 0 & \dots & \dots \\
& & & 0.800 & 1.245 & 7.71 & 10 & 0.333 & 0.180 \\
& & & 0.825 & 1.257& 8.09 &  10  & 0.209 & 0.076 \\
& & & 0.850 & 1.269 & 8.49 & 10  & 0.157 & 0.019 \\
& & & 0.875 & 1.282 & 8.89 & 10 & 0.141 & 0.010 \\
\tableline 
SASI & 0.0 & 0.1419 & 0.950 & 1.188 & 0.93 & 0 & \dots  & \dots \\
(S-Series) & & &0.975 & 1.195 & 0.97  & 1&  0.268 & \dots \\
& & &1.000 & 1.202 & 1.02 & 4 & 0.241  & \dots \\
& & &1.025 & 1.210 & 1.06 & 4 & 0.219 & \dots \\
& & &1.050 & 1.217 & 1.11 & 9 & 0.206 & \dots \\
& &  &1.075 & 1.225 & 1.15 & 9 & 0.239 & \dots \\
& & &1.100 & 1.233 & 1.20 & 9 & 0.243 & \dots\\
& &  &1.125 & 1.241 & 1.25 & 10 & 0.213 & 0.603 \\
\tableline
\end{tabular}
\tablecomments{Parameters of the initial conditions ($\varepsilon$, $A$, $B$, $r_S$, and $\chi$), and the number ($N_\mathrm{400}$), earliest time ($t_\mathrm{400,min}$), and time spread (${\Delta t}_\mathrm{400}$) of explosions obtained from 10 runs per value of heating parameter $B$ (where `explosion' here means the average shock radius reaching $400\,\mathrm{km}$).
Reference values: $\varepsilon_0 = G M / r_{S0}$, where $G$ is Newton's constant and $r_{S0} = 100~\mathrm{km}$; $A_0 = 145~{m_u}^{-1} {p_0}^{-3/2}~\mathrm{MeV}~\mathrm{s}^{-1}$, where $m_u$ is the atomic mass unit, $p_0 = (11 a_\gamma /12) (2~\mathrm{MeV})^4$, and $a_\gamma = 8.56\times 10^{31}~\mathrm{MeV}^{-3}\mathrm{cm}^{-3}$; and $B_0 = 160~m_u^{-1} (10^2~\mathrm{km})^2~\mathrm{MeV}~\mathrm{s}^{-1}$ \citep{Janka2001Conditions-for}.}
\end{center}
\end{table}

\begin{figure}
\centering
\includegraphics[width=0.32\textwidth]{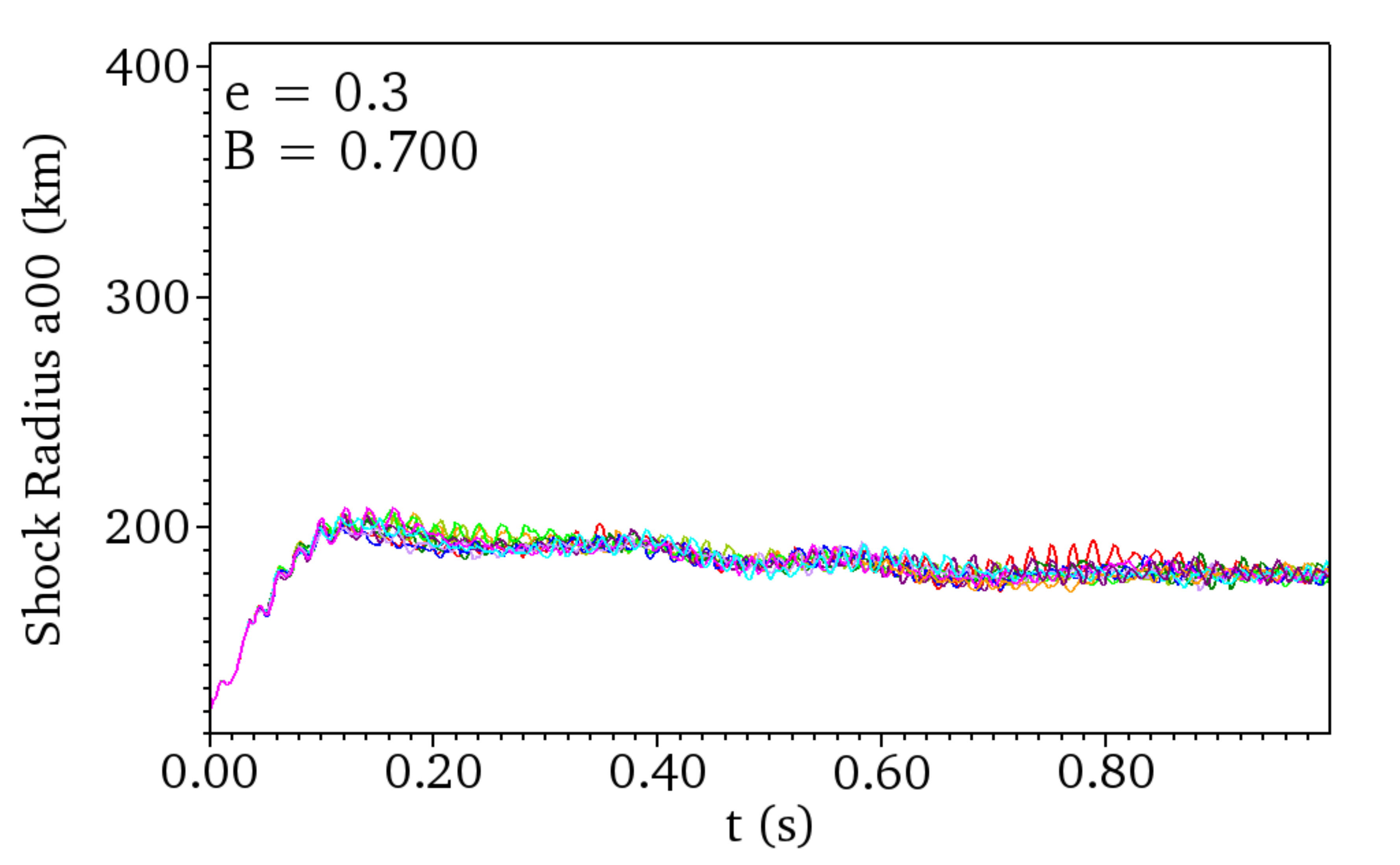}
\includegraphics[width=0.32\textwidth]{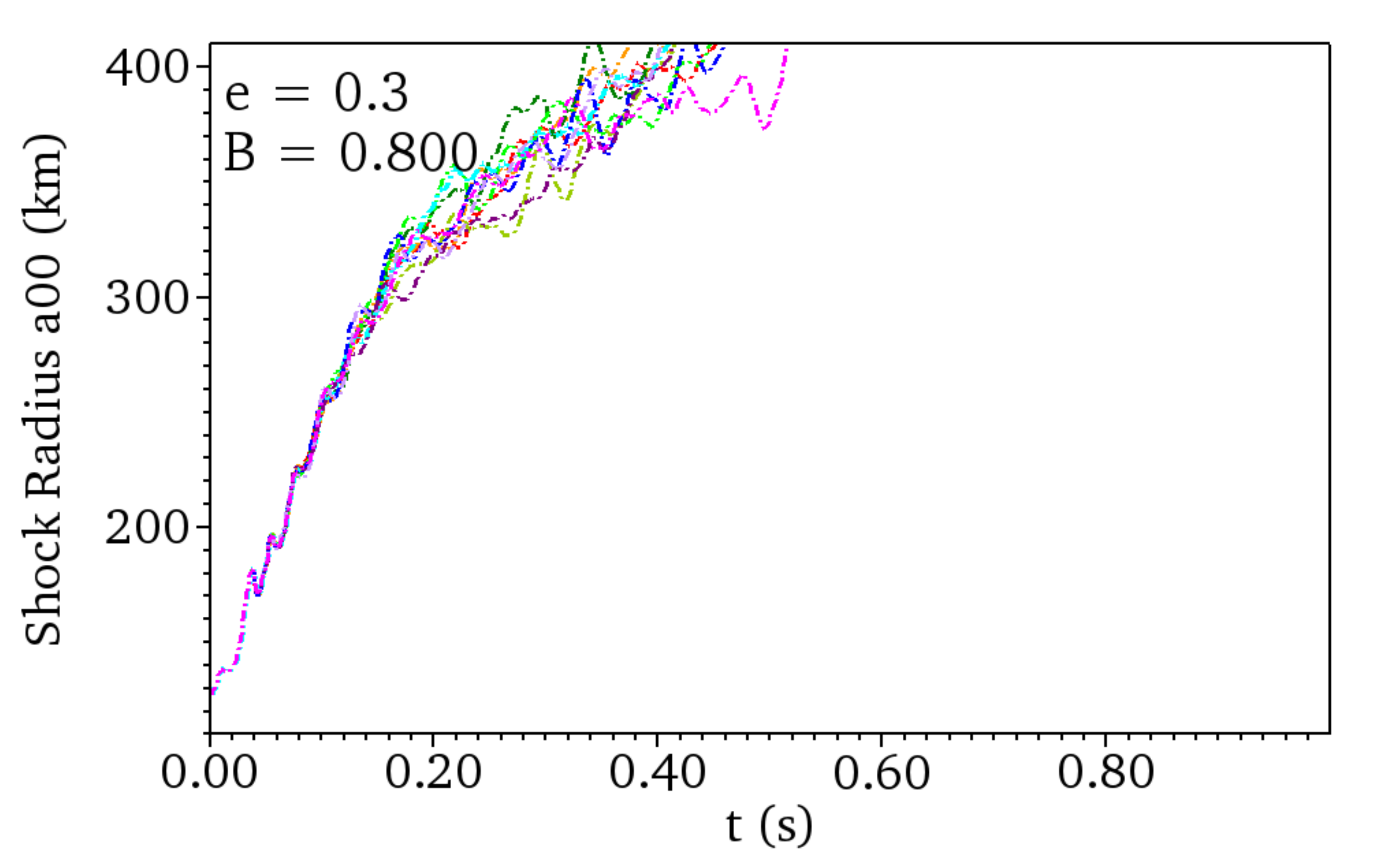}
\includegraphics[width=0.32\textwidth]{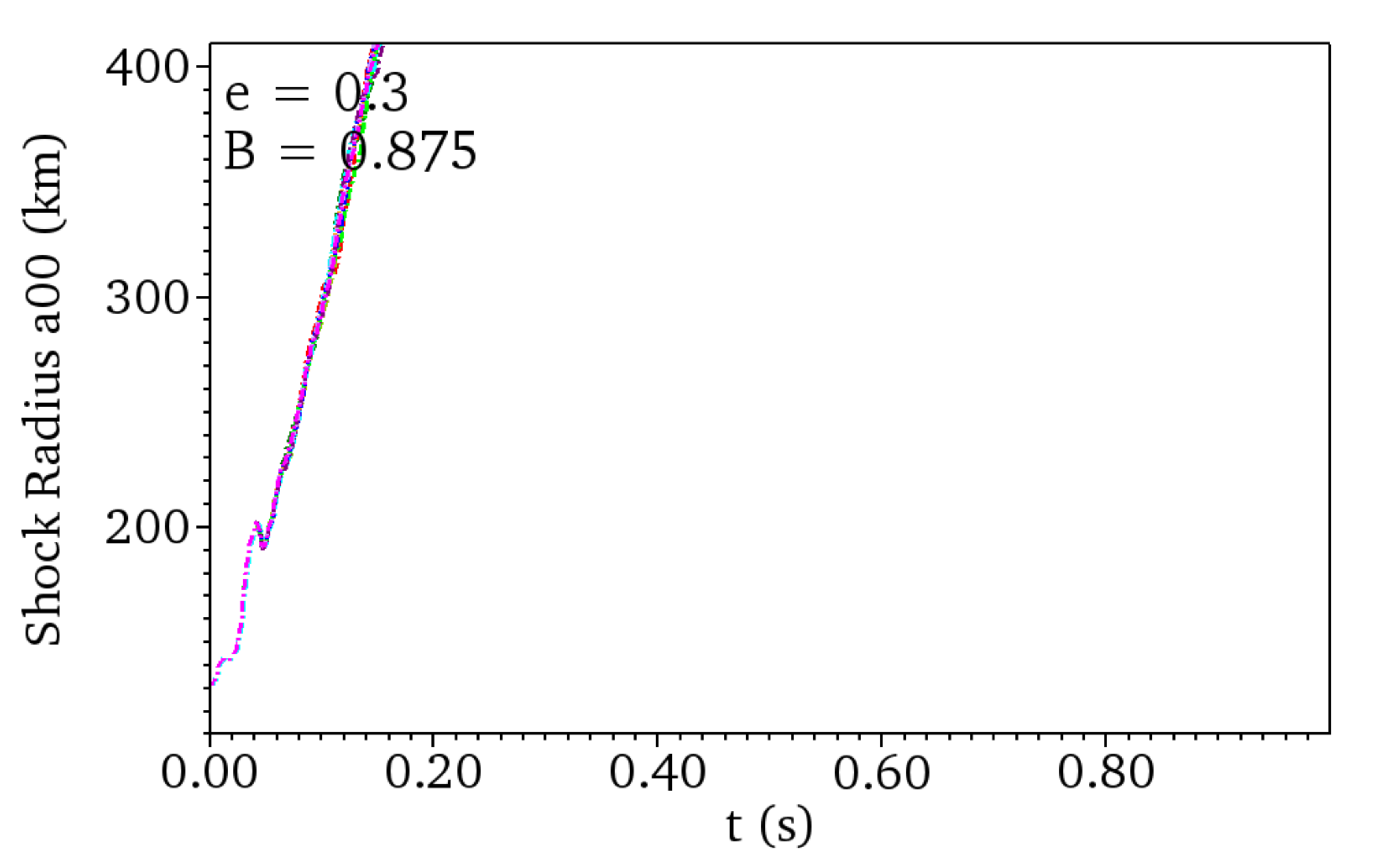}
\includegraphics[width=0.32\textwidth]{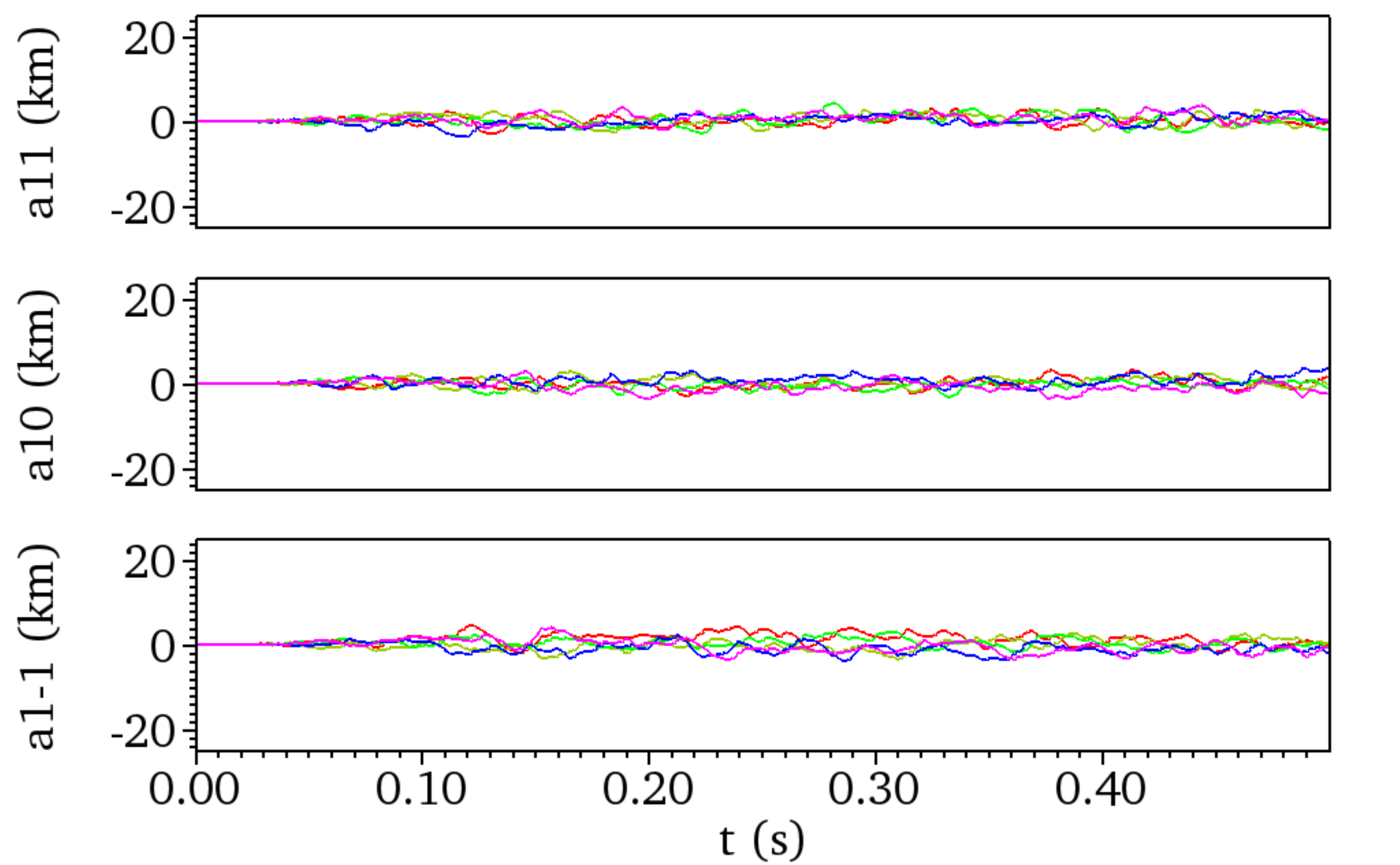}
\includegraphics[width=0.32\textwidth]{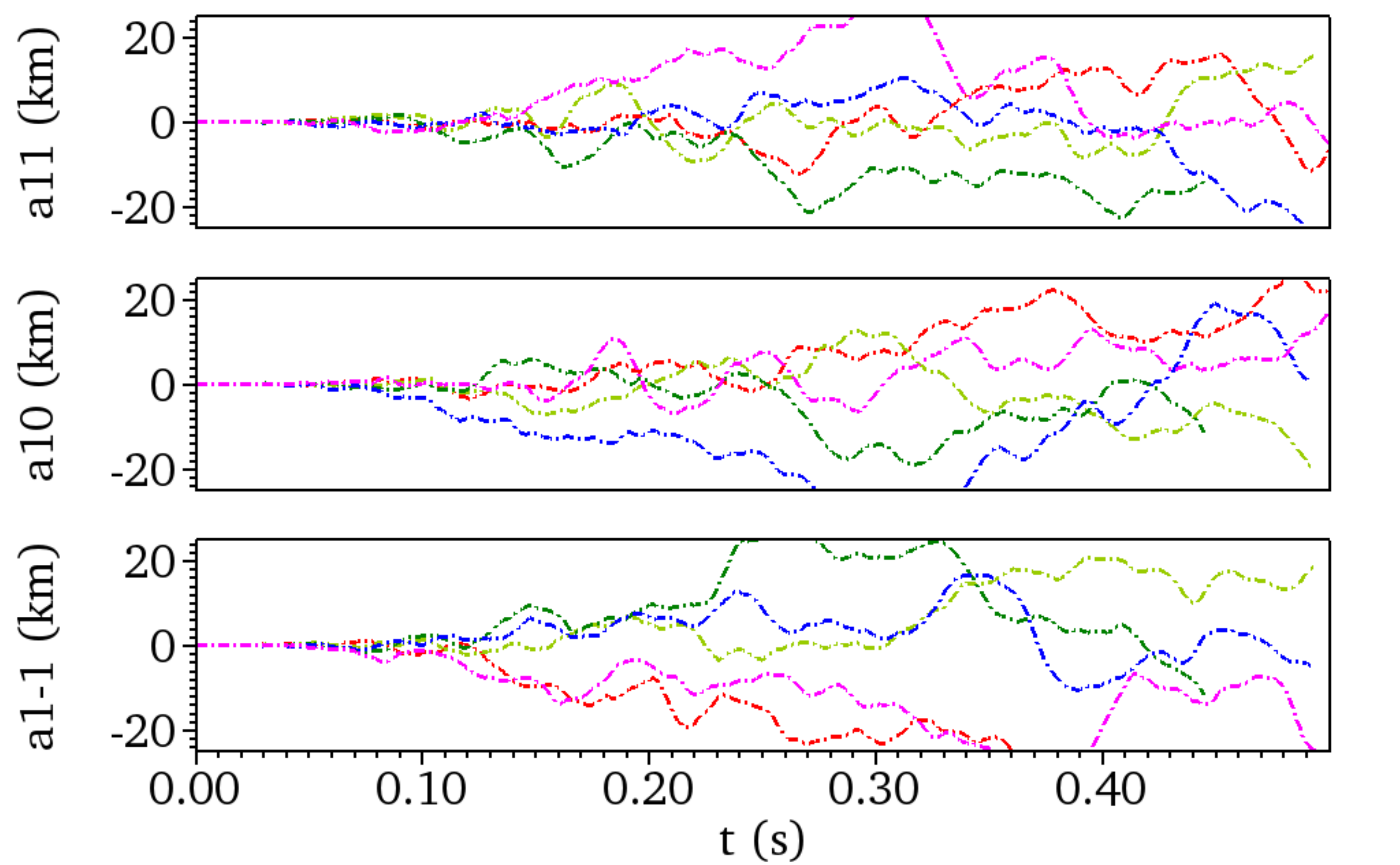}
\includegraphics[width=0.32\textwidth]{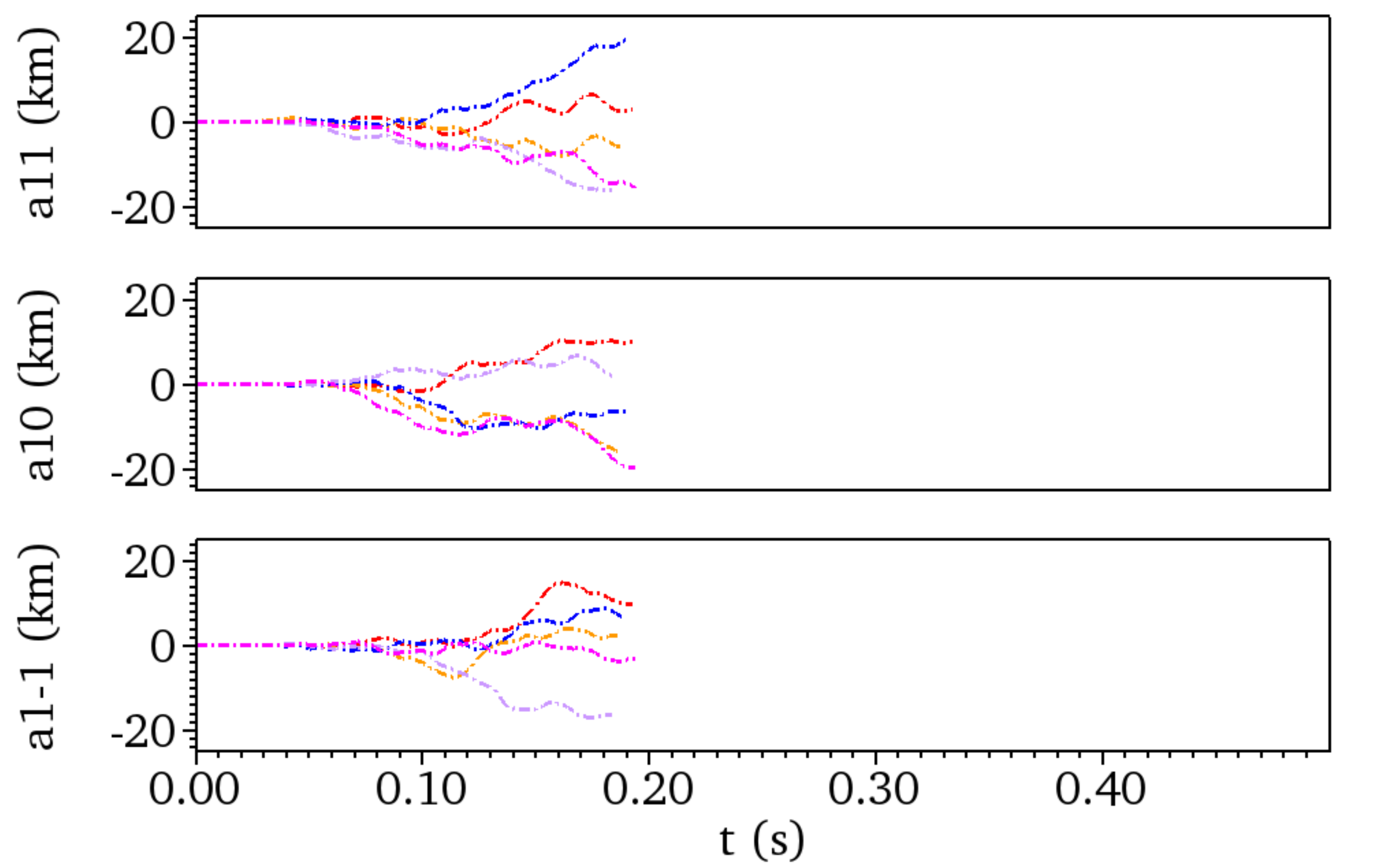}
\caption{C-Series shock radius decomposition coefficients for selected $B$ values. 
Solid lines are duds; dot-dashed lines are explosions.
Upper panels: average shock radius $a_{00}$ for all 10 runs for each value of $B$.
Lower panels: average $x$, $y$, and $z$ shock positions ($a_{11}$, $a_{1-1}$, $a_{10}$), with only half the runs and the first $0.5\,\mathrm{s}$ for clarity (the time scales differ from the upper panels).
\label{fig:C_Coefficients}}
\end{figure}

\begin{figure}
\centering
\includegraphics[width=0.32\textwidth]{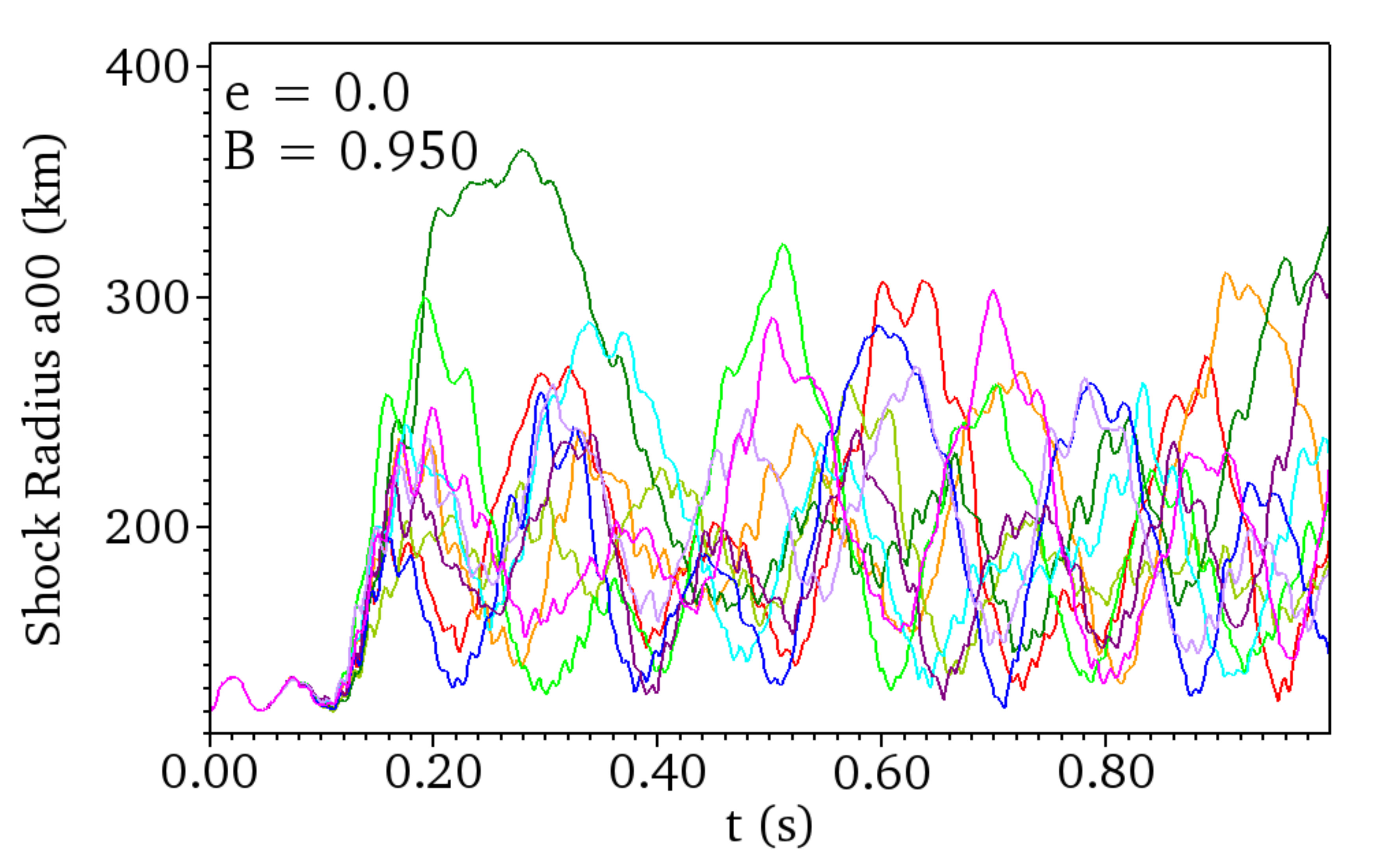}
\includegraphics[width=0.32\textwidth]{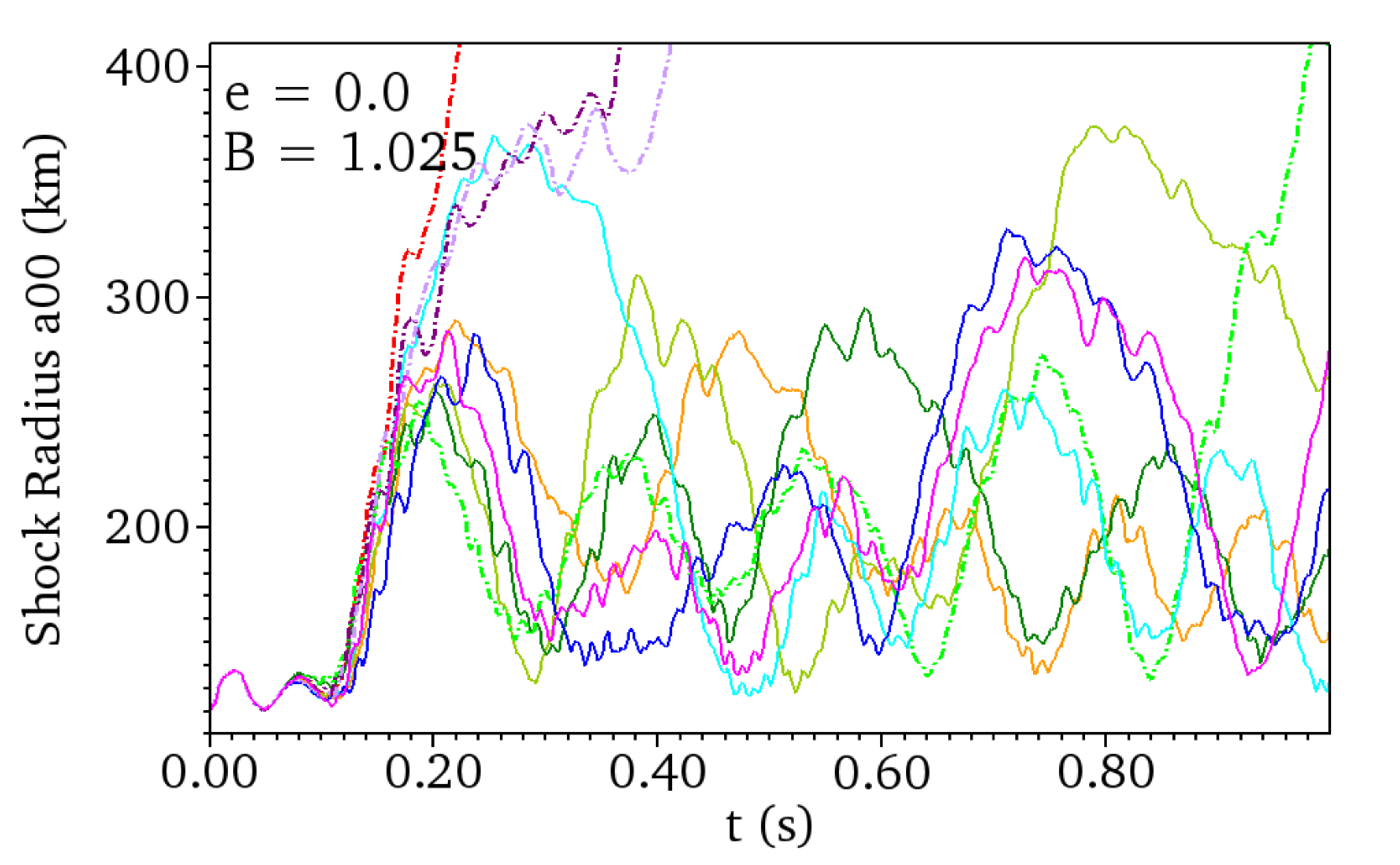}
\includegraphics[width=0.32\textwidth]{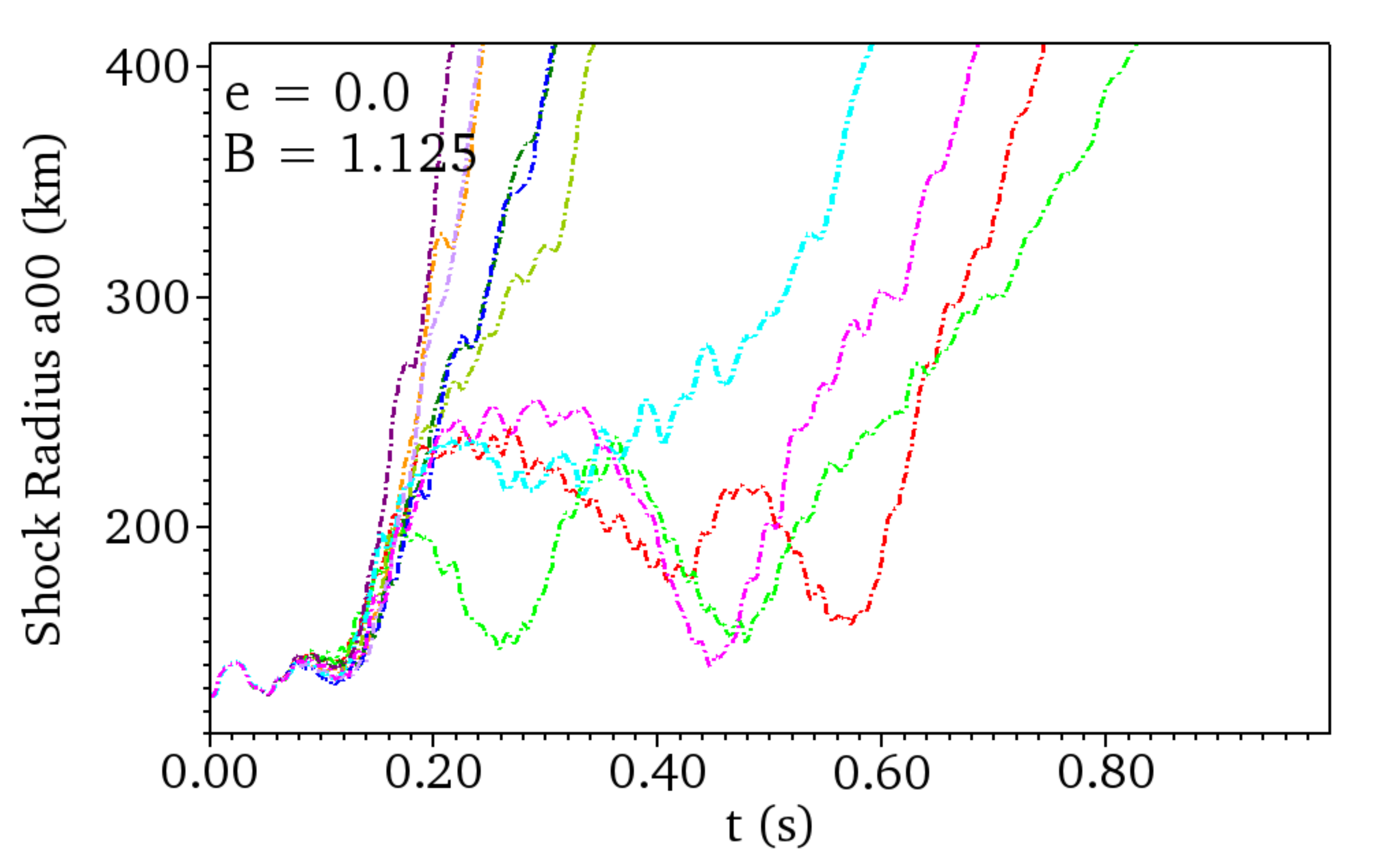}
\includegraphics[width=0.32\textwidth]{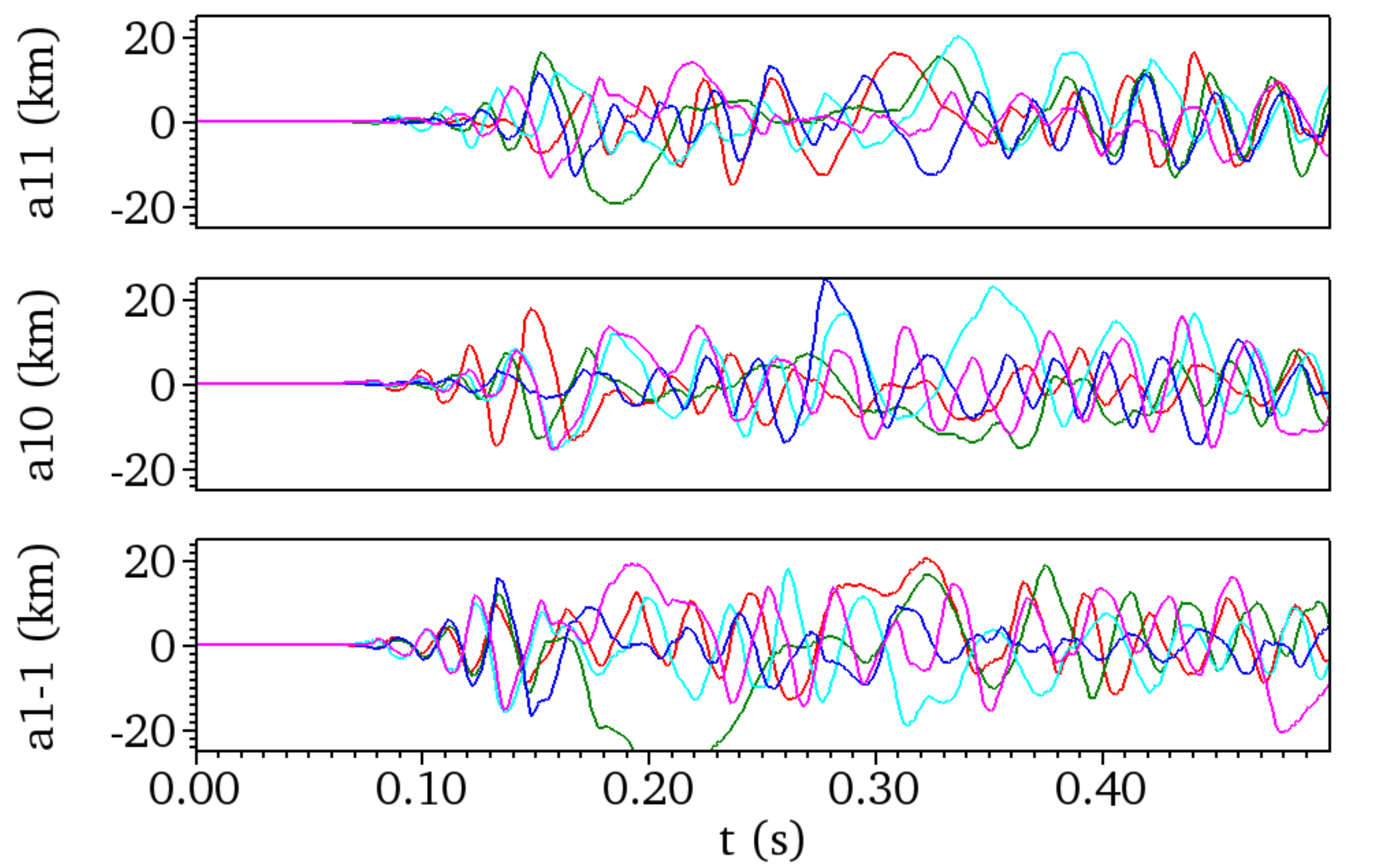}
\includegraphics[width=0.32\textwidth]{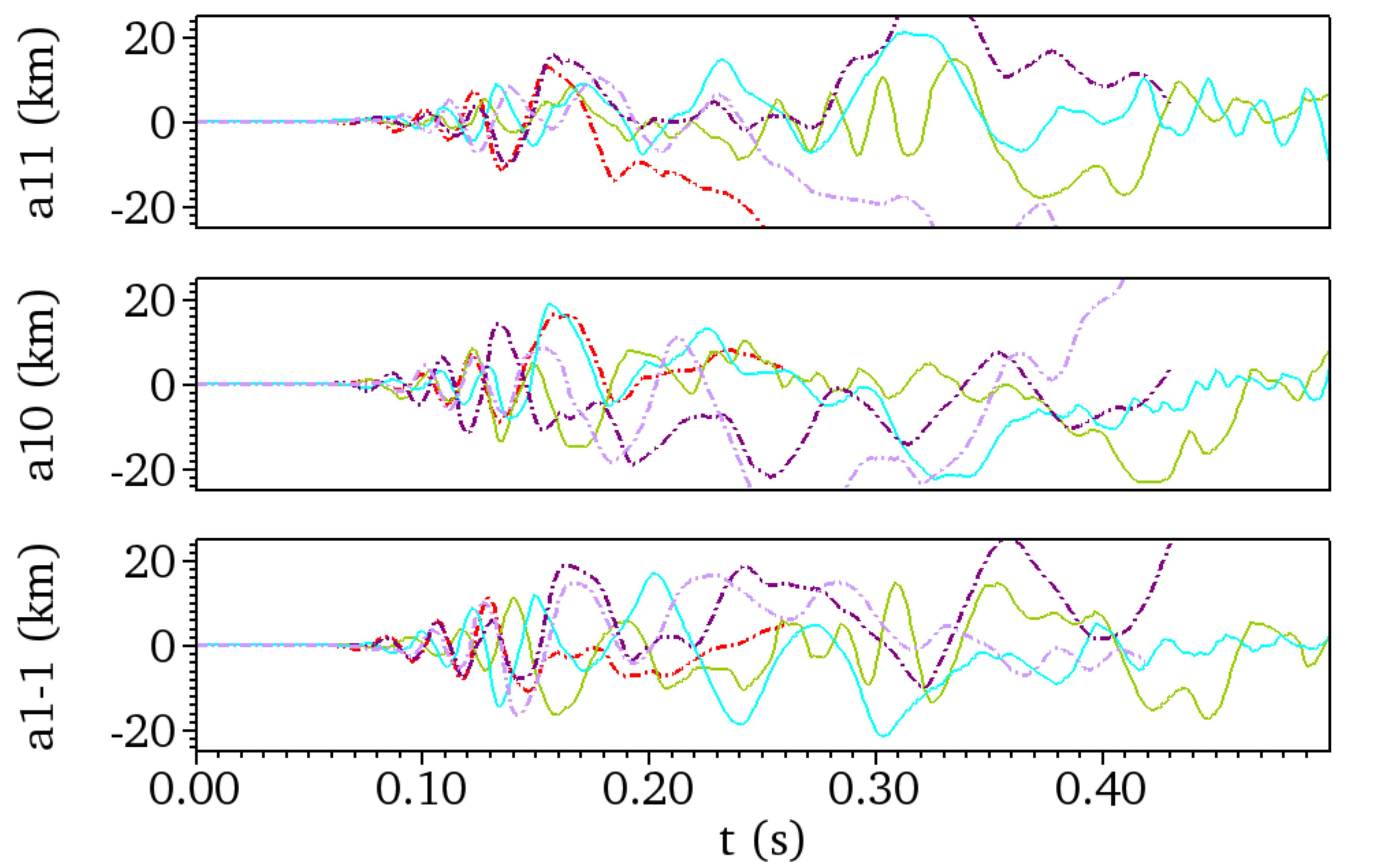}
\includegraphics[width=0.32\textwidth]{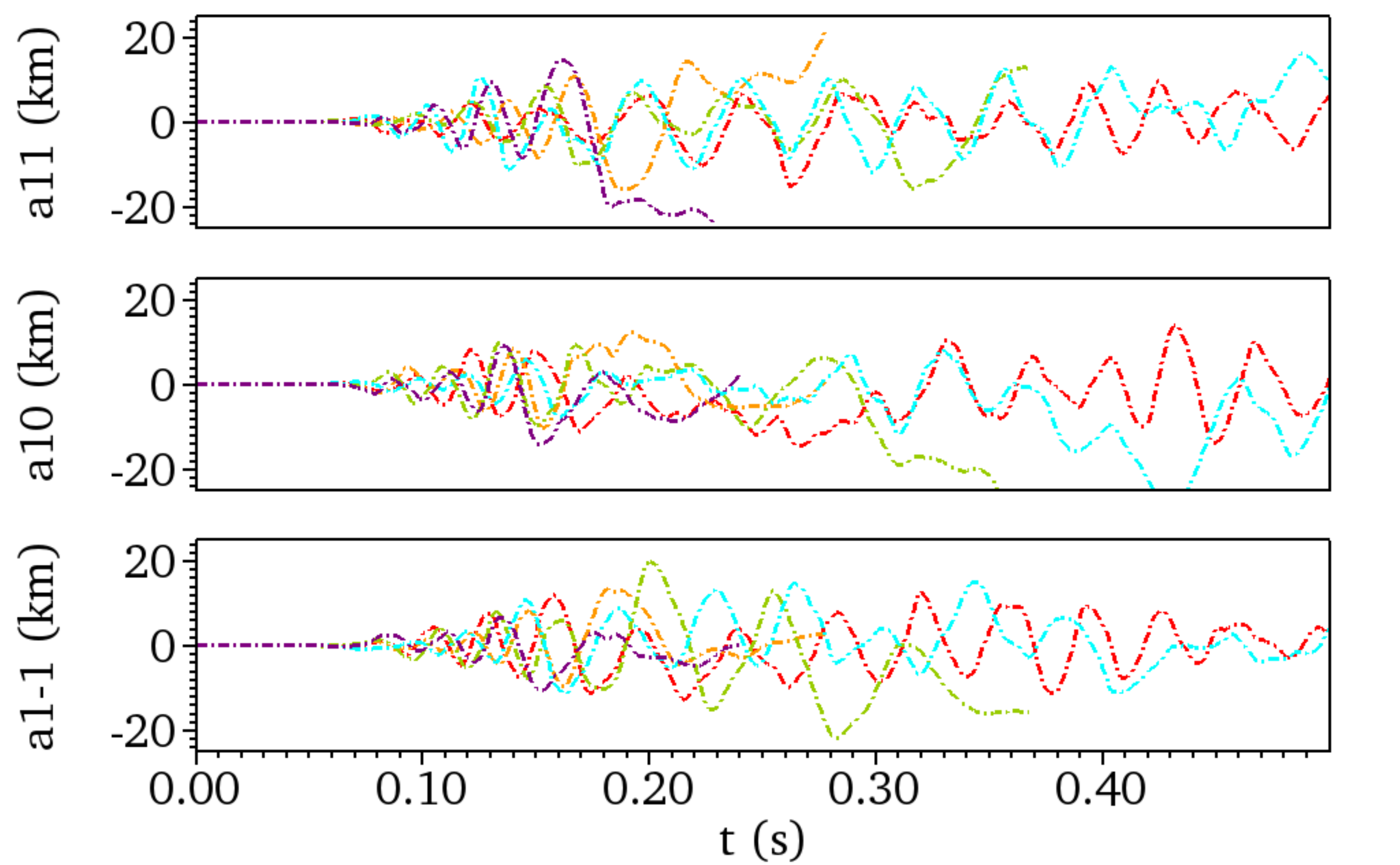}
\caption{
As in Figure~\ref{fig:C_Coefficients}, for the S-series.
\label{fig:S_Coefficients}}
\end{figure}

We summarize our simulations in Table~\ref{tab:SeriesParameters}, and more selectively in Figures~\ref{fig:C_Coefficients} and \ref{fig:S_Coefficients}. 
For each series, we run 10 simulations for each value of $B$ in Table~\ref{tab:SeriesParameters}, which summarizes some parameters of the initial conditions and statistics of the outcomes.
Figures~\ref{fig:C_Coefficients} and \ref{fig:S_Coefficients} survey the differing shock behavior of each series as a function of heating parameter:
for selected values of $B$ below, at, and above `threshold', the $\ell = 0, 1$ real spherical harmonic coefficients $a_{\ell m}$ of the shock surface are displayed for several simulations,
normalized such that $a_{00}$ is the average shock radius, and $a_{11}$, $a_{1-1}$, and $a_{10}$ are the average $x$, $y$, and $z$ positions of the shock \citep{Burrows2012An-Investigatio}.

For the C-series, the number of explosions $N_{400}$ as a function of $B$ tabulated in Table~\ref{tab:SeriesParameters} comports with the paradigmatic notion of a `critical' luminosity: $N_{400}$  jumps from 0 to 10 between values of $B$ separated by only $\sim 3\%$.
Mild stochasticity is evident in the spread of explosion times $(\Delta t)_\mathrm{400}$ in the C-series runs, which decreases with increasing $B$, as does
the earliest explosion time $t_\mathrm{400,min}$.

However, the critical luminosity concept is challenged by the more extreme stochasticity of the S-series, which extends beyond explosions times to whether an explosion occurs at all (at least within our $1\,\mathrm{s}$ simulation time\footnote{With our stationary luminosity and accretion rate, the concept of a critical luminosity might be recovered if the simulations ran indefinitely. But in nature, luminosity declines with time, leaving a finite time for explosion to occur. Thus, if SASI-dominated explosions occur in nature, the stochasticity of outcomes reported here may plausibly hold.}):
 this `threshold' is smeared out, with $N_{400}$ in Table~\ref{tab:SeriesParameters} increasing from 0 to 10 over a $\sim20\%$ range of $B$.
The spread of explosion times is only given for the largest $B$ value, the only one for which all 10 runs explode. 
This spread $(\Delta t)_\mathrm{400} = 0.603\,\mathrm{s}$ is significantly larger than in the C-series. 
Unlike the tabulated values of $t_\mathrm{400,min}$ for the C-series, there is no clear trend of earliest explosion time with increasing $B$; apparently, if an explosion can happen at all, it has a decent chance of occurring in the initial expansion, at roughly the same time for all tabulated $B$.

Figures~\ref{fig:C_Coefficients} and \ref{fig:S_Coefficients} reinforce the outcomes in Table~\ref{tab:SeriesParameters}.
Whereas the C-series runs in Figure~\ref{fig:C_Coefficients} show either steady shock expansion (explosions) or a stationary radius (duds), the strong influence of the SASI yields large, cyclic excursions on roughly the SASI growth timescale in the upper panels of Figure~\ref{fig:S_Coefficients}.
Again, the explosion threshold is not sharp for the S-series: the upper middle panel of Figure~\ref{fig:S_Coefficients} shows some runs exploding, and others not, within the time frame of the simulation.
Note also a telltale signature of the SASI in the lower panels of Figure~\ref{fig:S_Coefficients}---the oscillatory dipole coefficients.
These grow more irregular and episodic when convection develops and interrupts their global coherence.  
Such oscillations are absent from the lower panels of Figure~\ref{fig:C_Coefficients}; dipole coefficients in these convection-dominated models grow in a non-oscillatory manner as explosions take off and bubbles merge into dominant directions of expansion. 

\begin{figure}
\centering
\includegraphics[width=0.49\textwidth]{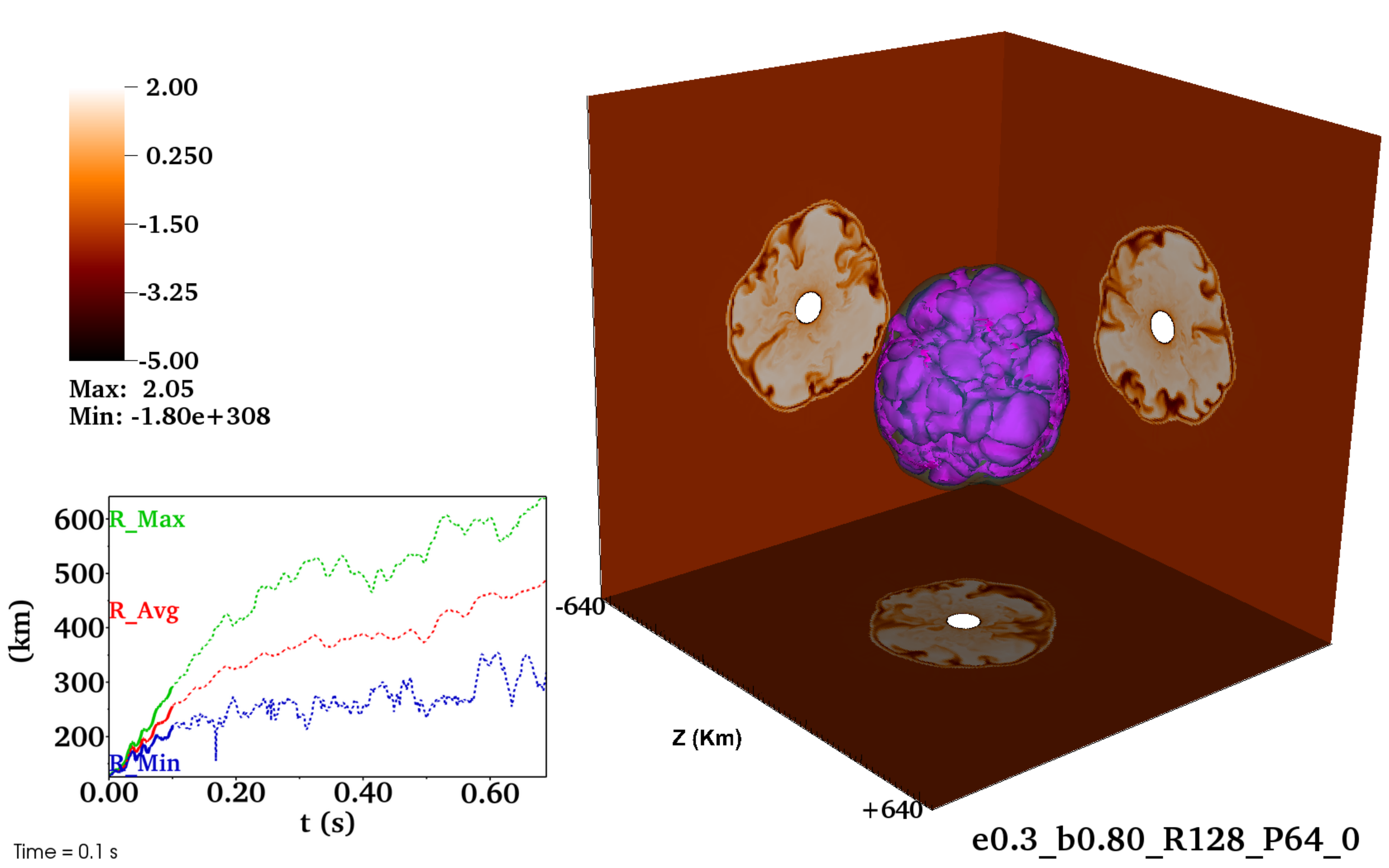}
\includegraphics[width=0.49\textwidth]{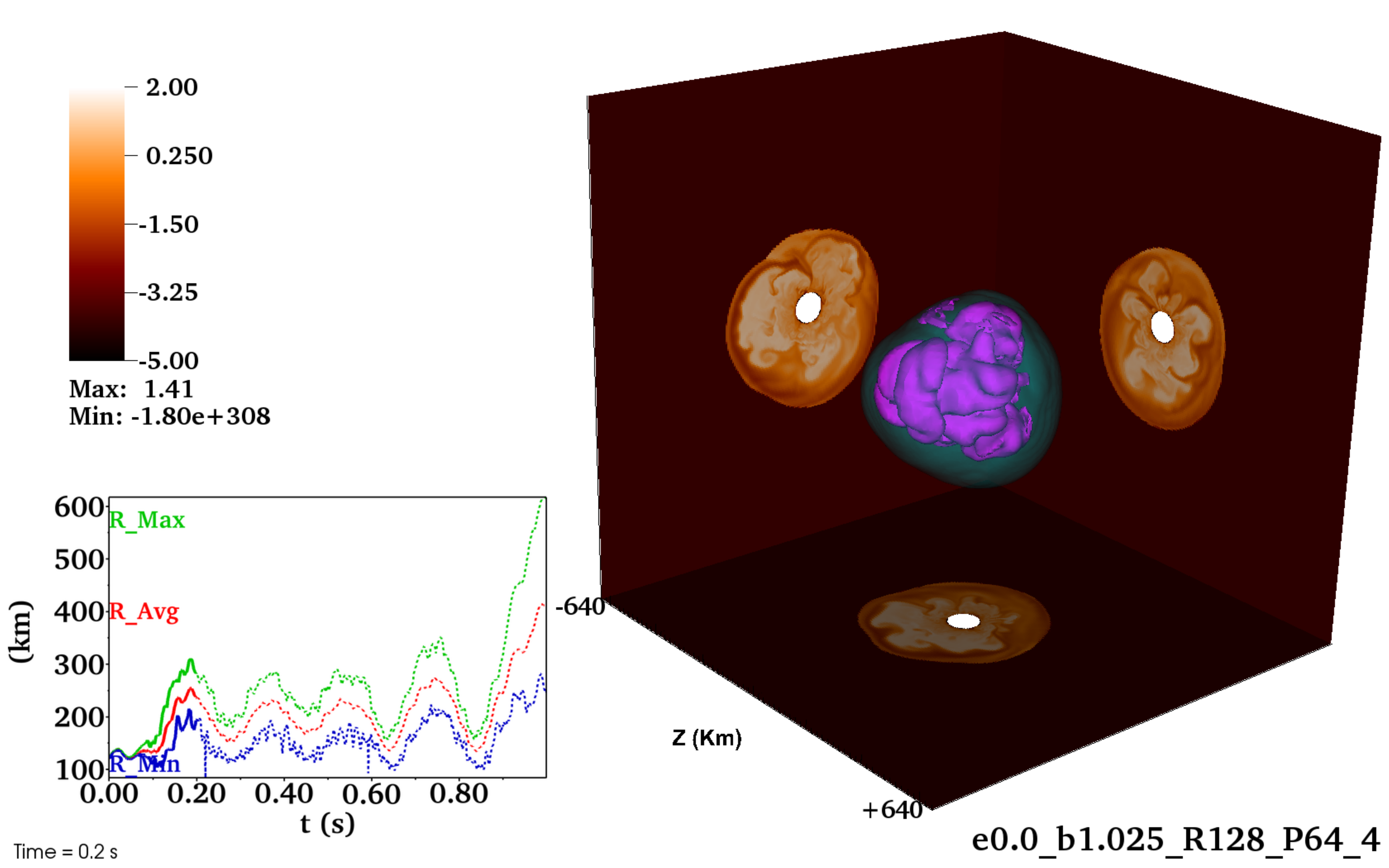}
\includegraphics[width=0.49\textwidth]{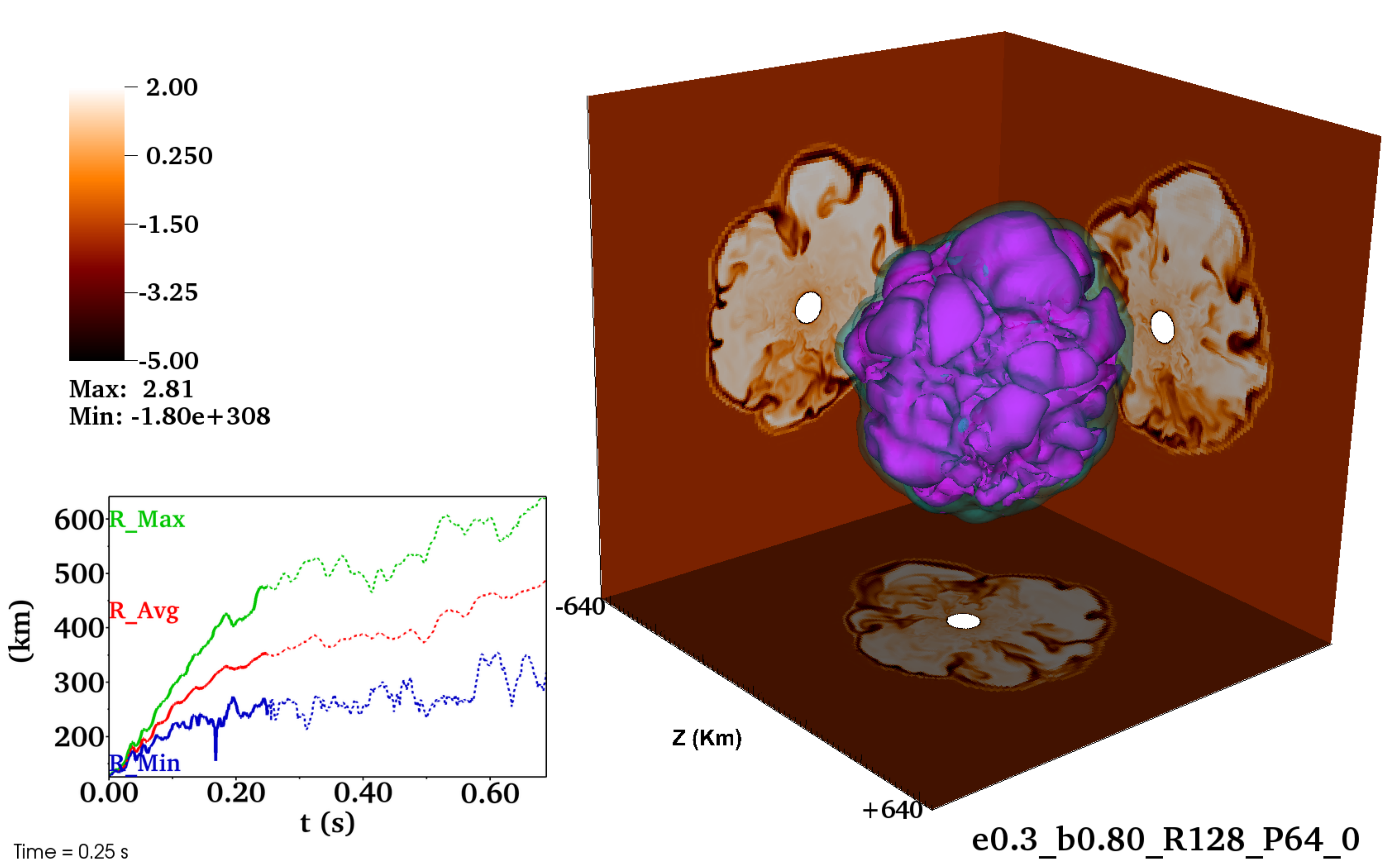}
\includegraphics[width=0.49\textwidth]{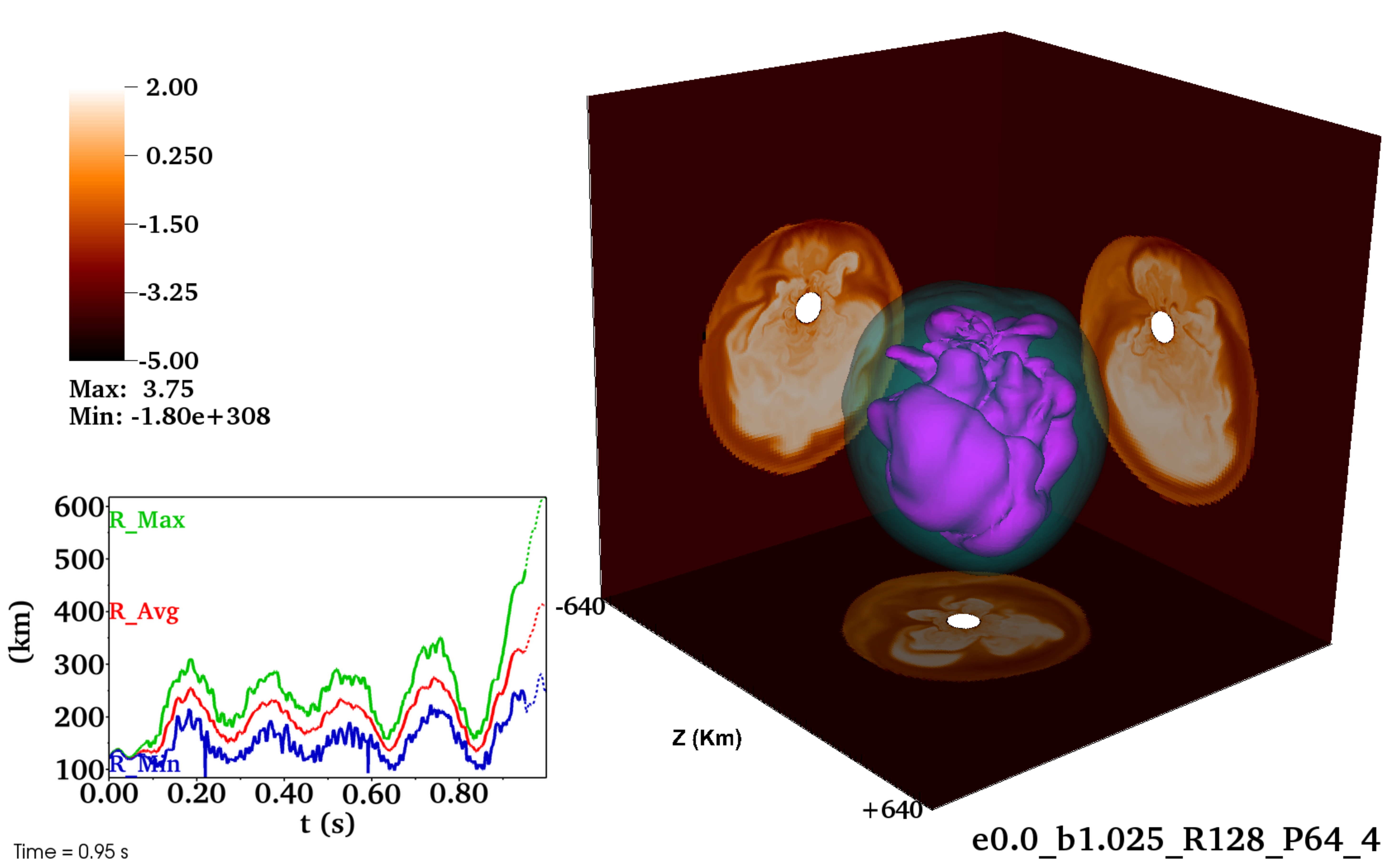}
\caption{Left panels: Sample convection-dominated simulation with $B=0.800\, B_0$ at $t=0.1\,\mathrm{s}$, and at $t=0.25\,\mathrm{s}$ as the explosion takes off. Right panels: Sample SASI-dominated simulation with $B=1.025\, B_0$ at $t=0.2\, \mathrm{s}$, and at $t=0.95\, \mathrm{s}$ as the explosion takes off.
Surface plots of two entropy values show the shock (cyan) and some structure in the post-shock region (purple).
Entropy cross sections through the origin are projected onto the back walls.
The average, maximum, and minimum shock radii are shown in the lower left of each panel, with the solid and dotted portions of these traces indicating times before and after the moment shown.
For each simulation, the zero of entropy is taken to be the immediate post-shock value at $t = 0$.
\label{fig:Selected}}
\end{figure}

Figure~\ref{fig:Selected} shows sample C-series and S-series runs.
In the exploding convective model (left panels), 
Rayleigh-Taylor plumes, driven by a negative entropy gradient generated by neutrino heating, grow from small scales and merge into larger bubbles, which become sufficiently buoyant and persistent to expand the shock to our outer boundary. 
In the SASI-dominated example (right-panels),
a hint of spiral behavior manifests in the upper left cross section of the upper right panel, with a plunging stream of low-density material around 11 o'clock separating counterrotating flows;
but the behavior is not pure SASI, as smaller-scale Rayleigh-Taylor plumes are also present.

Animations of S-series runs suggest that the long-time-scale, quasi-periodic excursions of the shock radius result from the palpable influence of both the SASI and neutrino heating. 
At small shock radii, SASI spiral waves appear. 
These increase the shock radius and generate entropy gradients to a point favorable to convection.
Even without neutrino heating, the Rayleigh-Taylor instability is a SASI saturation mechanism \citep{Guilet2010The-Saturation-};
here, neutrino heating takes buoyancy beyond \textit{saturation} of the SASI to \textit{interruption} of the SASI.
When convective bubbles are sufficiently persistent to interrupt the global SASI oscillation, but not sufficiently buoyant to drive runaway shock expansion, the shock radius collapses and the cycle begins anew.
In a final cycle towards the end of the run in the right panels of Figure~\ref{fig:Selected}, the actions of the SASI and convection coincide and reinforce each other to the point that a large, buoyant bubble drives runaway expansion. 

\begin{figure}
\centering
\includegraphics[width=0.5\textwidth]{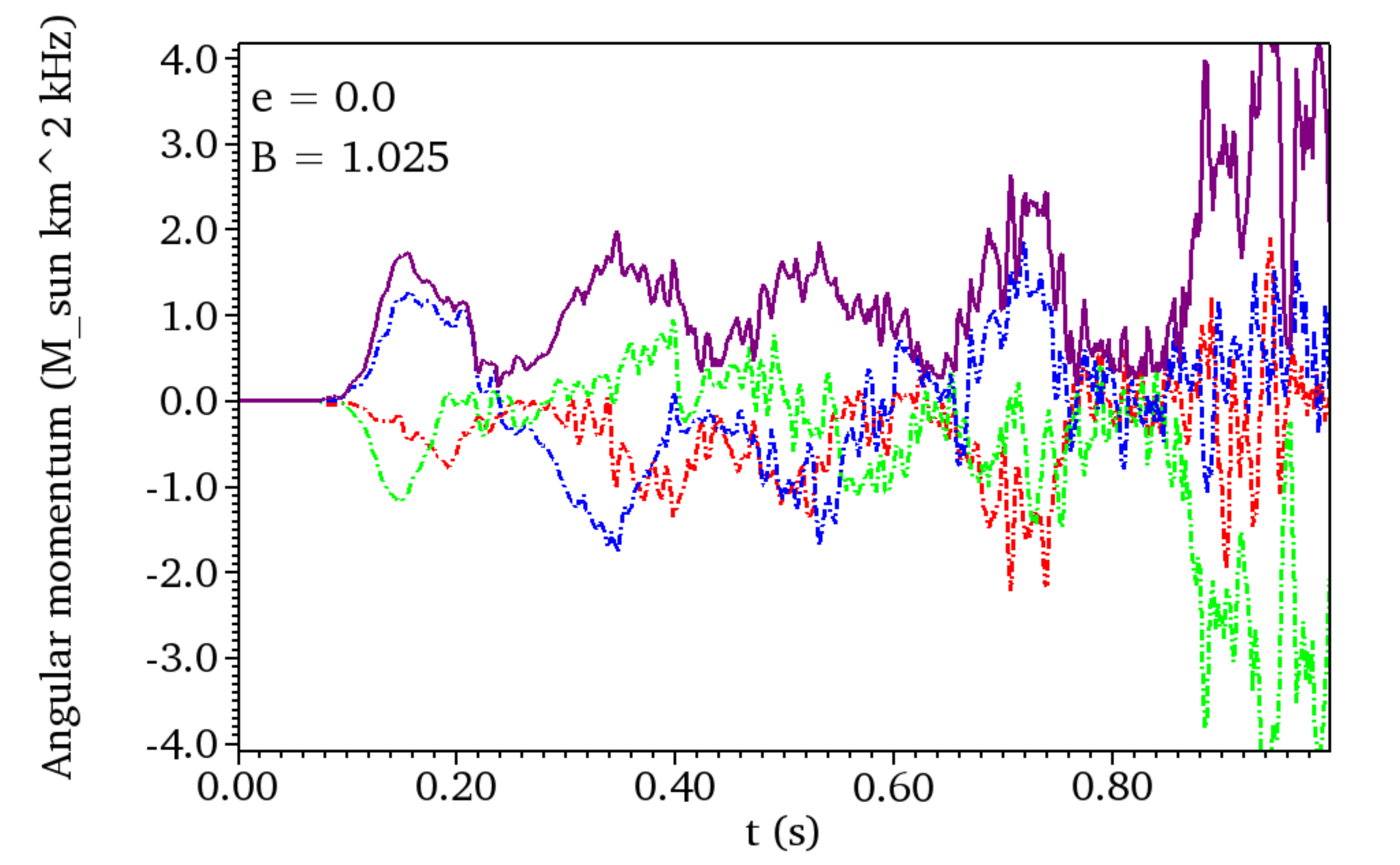}
\includegraphics[width=0.5\textwidth]{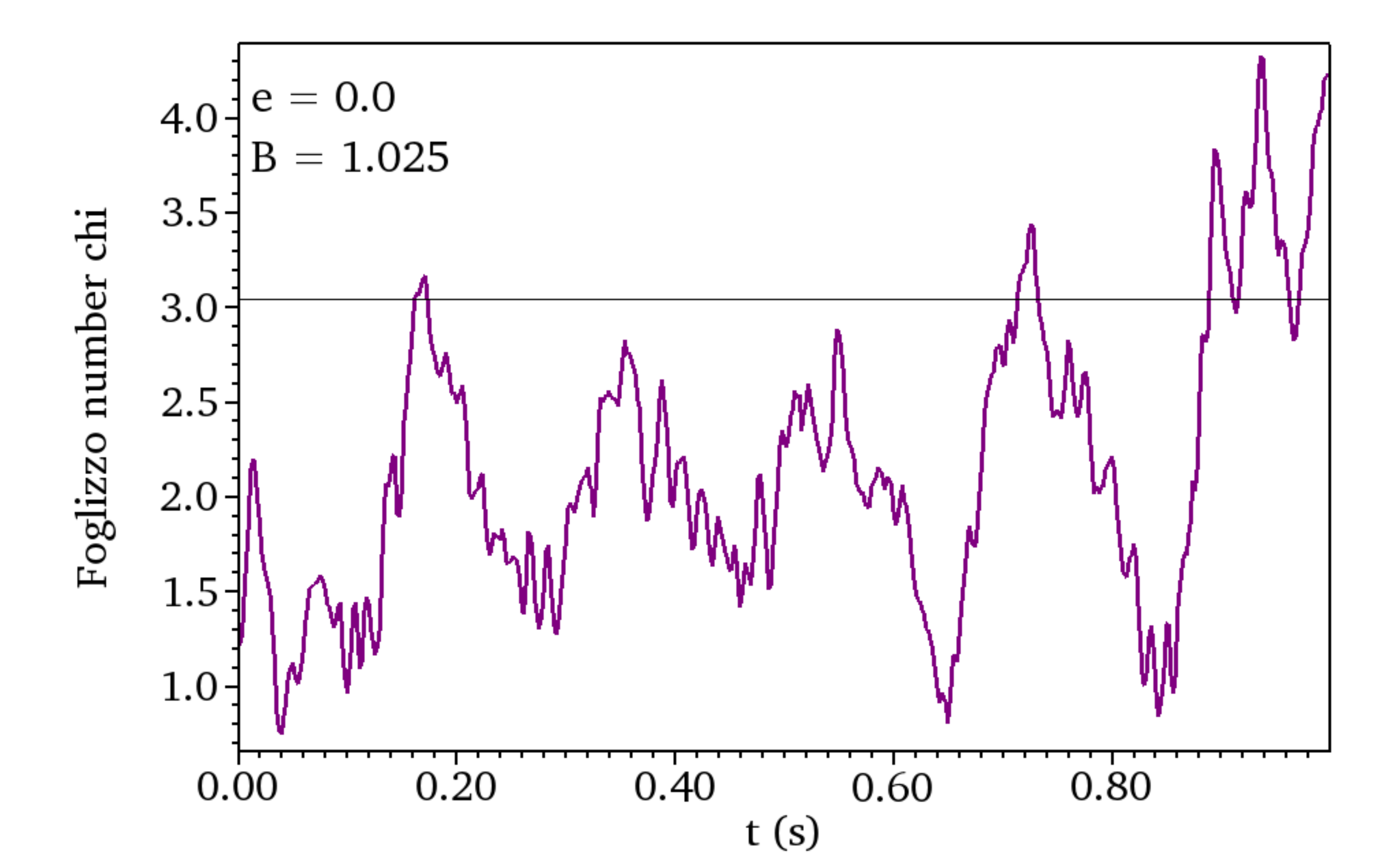}
\includegraphics[width=0.5\textwidth]{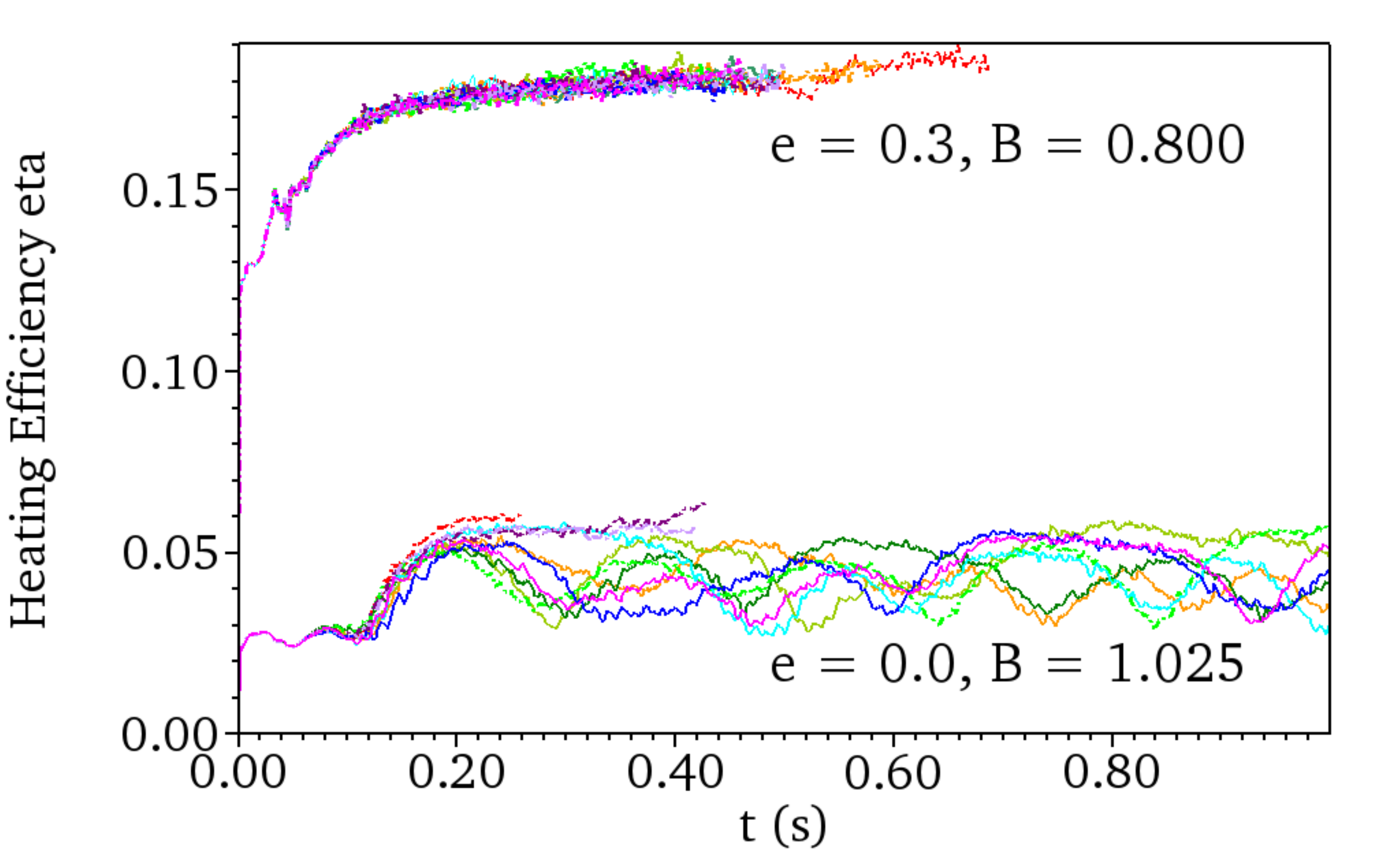}
\caption{Upper and middle panels: Angular momentum magnitude (solid) and components (dot-dashed), and Foglizzo number $\chi$, of the SASI-dominated run shown in the right panels of Figure~\ref{fig:Selected}.
Lower panel: Heating efficiency $\eta$ of convection-dominated runs with $B = 0.800\, B_0$ (upper curves) and SASI-dominated runs with $B = 1.025\, B_0$ (lower curves).
\label{fig:S_ChiEta}}
\end{figure}

This interpretation is supported by the time series of SASI-induced angular momentum $\mathbf{L}$, and $\chi$, in the upper and middle panels of Figure~\ref{fig:S_ChiEta}.
The radius changes in the lower left corners of the right panels of Figure~\ref{fig:Selected} are manifestly reflected in $|\mathbf{L}|$ and $\chi$.
Strikingly, the $\chi$ peaks are close to the critical value $\chi \simeq 3$ for convective activity to persist.
The shock expansion occasioned by the SASI gives convection a chance to take hold; but if runaway expansion does not immediately result, the shock recedes to a point that the spiral SASI is renewed with a stochastic reorientation, as indicated by the $\mathbf{L}$ component traces.

Finally, we compare the heating efficiencies in C-series and S-series runs near threshold, which suggests that SASI-dominated systems are significantly `easier' to explode.
Computed from spherically averaged radial profiles, the heating efficiency $\eta$ is the integrated net heating rate in the `gain region' (where the net heating is positive) divided by the sum of the luminosities of the neutrino species responsible for heating [($L_{\nu_e} + L_{{\bar\nu}_e}$, which is related to our heating parameter $B$ \citep{Janka2001Conditions-for}].
The lower panel of Figure~\ref{fig:S_ChiEta} shows $\eta$ for the C-series runs just above threshold ($B = 0.800\, B_0$) and for the S-series runs towards the middle of the `smeared out' threshold region ($B = 1.025\, B_0$). 
Only 4 of these SASI-dominated runs explode, but the heating efficiencies of these exploding runs are not drastically different from the 6 duds.
For explosions, we see that heating efficiencies 3 to 4 times higher are needed for convection-dominated runs than for SASI-dominated runs.

The lower efficiencies needed for SASI-dominated systems to explode imply that they are more `forgiving,' in that they require less from the neutrino heating.
This is physically plausible.
In convection-dominated systems, the neutrino heating does double duty: in addition to inflating bubbles to runaway buoyancy, neutrino-driven convection must be initially strong enough to expand the shock against gravity and ram pressure to an extent that makes such inflation possible;
whereas in SASI-dominated systems, the SASI initially does much of the `heavy lifting' in terms of initial shock expansion and generation of entropy gradients, such that all that is required of the neutrino heating is the inflation of bubbles in conditions that are already more favorable.

\section{Conclusion}

We ran 160 simulations with a highly simplified model of the post-bounce supernova environment in 3D.
This simplified model allows control over the nature of the dominant multidimensional instability affecting shock evolution.

Do core-collapse supernovae exhibit qualitatively significant sensitive dependence on initial conditions? \citet{Wongwathanarat2013Three-dimension}, \citet{Handy2014Toward-Connecti}, \citet{Takiwaki2014A-Comparison-of}, and \citet{Iwakami2014Parametric-Stud}---with more physics, though in most cases still parametrized, and with relative handfuls of 3D realizations---find quantitative spreads in explosion time  and energy, along with larger variations in ejecta morphology, neutron star spins and kick velocities, and distributions of nucleosynthesis products. 
We also see quantitative spreads in explosion time in our convection-dominated models, which decrease with increasing luminosity.

However, we find that stochasticity extends to the qualitative outcome of explosion vs. dud in the SASI-dominated case.
Out of 10 simulations for each value of heating parameter $B$ listed for the SASI-dominated models in Table~\ref{tab:SeriesParameters}, the number of explosions increases from 0 to 10 over a $\sim 20 \%$ range in luminosity.
This stochasticity in qualitative outcome arises from the strong interplay of two instabilities. 
The SASI is enabled first, increasing the shock radius to a point that allows convection to begin.
Subsequent evolution depends on whether neutrino heating drives runaway bubble inflation at this key moment.
If not, the shock recedes and the cycle repeats with the SASI again taking over.
This long-period cyclic interaction has been glimpsed before---see \citet{Iwakami2014Parametric-Stud} Figure~2, \citet{Fernandez2015Three-dimension} Figure~3, and some self-consistent models \citep{Hanke2013SASI-Activity-i,Tamborra2014Self-sustained-}.
But a larger number of realizations and finer sampling of heating strength have here revealed that this cyclic interaction leads to more qualitative stochasticity than in the convection-dominated case.
The increasing probability of explosion with increasing heating parameter in the SASI-dominated case undermines the paradigmatic notion of a critical luminosity for explosion \citep{Burrows1993A-Theory-of-Sup}, or at least modifies it to a smeared-out `threshold region.'\footnote{The increase of the SASI-dominated luminosity threshold with higher resolution reported by \citet{Fernandez2015Three-dimension}, while physically plausible, requires reassessment with more realizations. Such a study would also indicate how the magnitude of threshold smearing---roughly $20\%$ in the present study---may depend on resolution.}

Is the SASI a potentially important contributor to supernova explosions?
With more complete physics, it has tended to appear in cases that do not explode,
leading some to believe that the SASI is a curiosity that does not directly contribute to explosions in nature. 
However, our results on the heating efficiencies of convection-dominated vs. SASI-dominated explosions motivate an open mind, particularly given the historical lack of robustness of the delayed neutrino-driven mechanism.
At first blush, the lower efficiencies of the SASI-dominated case in the lower panel of Figure~\ref{fig:S_ChiEta} may seem \textit{less} conducive to explosions, for we usually hear about multidimensional effects \textit{increasing} heating efficiency.
That may be true of convection. 
But our intuition about neutrino heating efficiency can be turned on its head: if a model can explode with lower efficiency, it requires less from the neutrino heating, making the neutrino-driven mechanism more robust.\footnote{See  \citet{Couch2015The-Role-of-Tur} on this `inverted' perspective in connection with turbulence.}

We have artificially induced the SASI by `turning off' our mock-up of nuclear dissociation,\footnote{Another artificial means of investigating explosions---increasing neutrino luminosities beyond nominally physical values---tends to suppress the SASI, perhaps contributing to prejudice against it.} but there are also physical possibilities for making the SASI more likely.
Several actively-studied aspects of progenitor structure affect the ease with which the SASI develops \citep{Hanke2013SASI-Activity-i,Couch2015The-Role-of-Tur,Muller2015Non-radial-inst}.
The equation of state at low density affects the shock jump conditions, and at high density affects the neutron star radius and speed of contraction.
Neutrino transport impacts the stalled shock radius and the Foglizzo number $\chi$; and note for instance that a change in neutrino opacities produced stronger SASI activity in a specific case \citep{Melson2015Neutrino-driven-II}. 
We should remain open to such possibilities conducive to the SASI in searching for a robust explosion mechanism, for nature may take advantage of its ability to do some of the `heavy lifting.'



\acknowledgments

We thank Rodrigo Fern\'andez for answering questions about his models.
We thank Eirik Endeve and Anthony Mezzacappa for discussions and ongoing collaboration.
This material is based upon work supported by the U.S. Department of Energy, Office of Science, Office of Advanced Scientific Computing Research and Office of Nuclear Physics.  
This research used resources of 
the Joint Institute for Computational Sciences at the University of Tennessee.

\def\cpc{Comp. Phys. Comm. } 


\clearpage

\end{document}